\documentclass[12pt]{article}
\usepackage{amsmath,amssymb}
\newtheorem{thm}{Theorem}[section]

\newtheorem{lem}[thm]{Lemma}

\newtheorem{assumption}[thm]{Assumption}

\newtheorem{definition}[thm]{Definition}

\newtheorem{example}[thm]{Example}

\newtheorem{remark}[thm]{Remark}
\newenvironment{rem}{\begin{remark}\rm}{\end{remark}}
\newtheorem{protocol}[thm]{Protocol}

\newcommand{\calL}{\mathcal{L}}

\newcommand{\F}{\mathbb{F}}

\newcommand{\Z}{\mathbb{Z}}

\newcommand{\lcm}{\mathrm{lcm}\,}
\newcommand{\eqr}[1]{~\mbox{$(${\rm \ref{#1}}$)$}}

\newcommand{\Section}[1]{\section{#1}\setcounter{equation}{0}}

\newcommand{\openbox}{\leavevmode
  \hbox to.77778em{%
    \hfil\vrule
  \vbox to.675em{\hrule width.6em\vfil\hrule}%
  \vrule\hfil}} 
\newcommand{\proofname}{Proof}

\newenvironment{proof}[1][\proofname]{\par\normalfont
  \trivlist\item[\hskip\labelsep\itshape #1:]\ignorespaces
  }{\hspace*{1cm}\hspace*{\fill}\openbox \medskip\endtrivlist}

\title{A Polynomial Description of the\\
  Rijndael Advanced Encryption Standard
\thanks{Supported in part by NSF grant DMS-00-72383.}}%
\date{April 30, 2002}%
\author{Joachim Rosenthal \\
  {\small Department of Mathematics\vspace{-2mm}}\\
  {\small University of Notre Dame\vspace{-2mm}}\\
  {\small Notre Dame, Indiana 46556, USA}\\
  {\small {\em e-mail:\/} Rosenthal.1@nd.edu}\vspace{-2mm}\\
  {\small http://www.nd.edu/\~{}rosen/ }}

\begin{document}\maketitle
\thispagestyle{empty}
\begin{abstract}
  The paper gives a polynomial description of the Rijndael
  Advanced Encryption Standard recently adopted by the National
  Institute of Standards and Technology. Special attention is
  given to the structure of the S-Box.\bigskip 


\end{abstract}

\Section{Introduction}                  \label{Sect:Int}

On November 26, 2001 the National Institute of Standards and
Technology (NIST) announced that the Rijndael encryption
algorithm becomes the Advance Encryption Standard.  The Rijndael
system will be a Federal Information Processing Standard (FIPS)
to be used by U.S. Government organizations (and others) to
protect sensitive information~\cite{ae01}. Detailed information
can be found at the website:\\
\centerline{http://csrc.nist.gov/encryption/aes/rijndael/}

The description~\cite{da99,da02b} supplied by Joan Daemen and Vincent
Rijmen, the inventors of the Rijndael encryption algorithm, is
very detailed and a reader new to the subject will probably need
some time to understand all steps in the algorithm.

In this paper we show how the whole algorithm can be quite
elegantly described through a sequence of algebraic manipulations
in a finite ring. We hope that this description will be helpful
in the proliferation of this new important standard.

We are aware of some attempts (e.g.~\cite{fe01,mu00}) where
authors tried to explore an algebraic description of the so
called `S-Box', the main non-linear part of the Rijndael system.
We are explaining in this paper why the `S-Box' can be described
through a sparse polynomial. There is however no attempt done to
explore this description further in order to find any weakness of
the system. We also derive the interpolation polynomial of the
inverse S-Box and we describe the cycle decomposition of the
S-Box.  The most detailed description of Rijndael can be found in
the new book~\cite{da02b}. This book gives many details on the
design philosophy and implementation aspects, something we do not
address in this paper.  During the preparation of this paper we
found the description of Rijndael as given in~\cite{tr02b}
useful. We want to thank U. Maurer for pointing us to an
algebraic description of Rijndael recently provided by
H.~W.~Lenstra~\cite{le02p}.

\Section{The Rijndael Algorithm}   \label{Sect:Dif}

Let $\Z_2=\{ 0,1\}$ be the binary field and consider the
irreducible polynomial
$$
\mu(z):=z^8+z^4+z^3+z+1\in \Z_2[z].
$$
Let $\F:=\Z_2[z]/<\mu(z)>=\mbox{GF}(256)$ be the Galois field of
$2^8$ elements and consider the ideal:
$$
I:= <x^4+1,y^4+1,\mu(z)>\subset \Z_2[x,y,z].
$$
We will describe the Rijndael algorithm through a sequence of
polynomial manipulations inside the finite ring
\begin{equation}
R:=\Z_2[x,y,z]/I=\F[x,y]/ <x^4+1,y^4+1>.
\end{equation}

The ring $R$ has simultaneously the structure of a finite
$\Z_2$-algebra and the structure of a finite $\F$-algebra as
above description makes it clear.  The monomials
$$
\left\{ x^iy^jz^k\mid 0\leq i,j\leq 3,\ 0\leq k\leq 7\right\}
$$
form a $\Z_2$-basis of the ring (algebra) $R$. In particular
$\dim_{\Z_2}R=128$, i.e.  $|R|=2^{128}$. Computations in the ring
$R$ can be done very efficiently. Addition in $R$ is done
componentwise and multiplication in $R$ is done through
multiplication in $\Z_2[x,y,z]$ followed by reduction modulo the
ideal $I$.
\begin{rem}
  One readily verifies that $x^4+1,y^4+1,\mu(z)$ forms a reduced
  Gr\"obner basis of the ideal $I$ which is also a
  zero-dimensional ideal. As a consequence the reduction modulo
  $I$ is very easy.  More details about finite dimensional
  algebras and zero dimensional ideals can be found
  in~\cite[Chapter~2]{co98}.
\end{rem}

Whenever $r\in R$ is an element we will define elements
$r_{i,j}\in\F$ and $r_j\in\F[x]/<x^4+1>$ through:
\begin{equation}                               \label{induced}
r=\sum_{i=0}^3\sum_{j=0}^3 r_{i,j}x^iy^j=
\sum_{j=0}^3\left( \sum_{i=0}^3  r_{i,j}x^i\right) y^j=
\sum_{j=0}^3 r_{j}y^j.
\end{equation}

On an abstract level a secret key crypto-system consist of a
message space $M$, a cipher space $C$ and a key space $K$
together with an encryption map
$$
\varepsilon : M\times K\longrightarrow C
$$
and a decryption map
$$
\delta : C\times K\longrightarrow M
$$
such that $\delta(\varepsilon(m,k),k)=m$ for all $m\in M$ and
$k\in K$. It should be computationally not feasible to compute
the secret key $k\in K$ from a sequence of plain-text/cipher-text
pairs $\left(m^{(t)},c^{(t)}=\varepsilon(m^{(t)} ,k)\right)$, $t=1,2,\ldots$.

In the Rijndael AES system one has the possibility to work with
secret keys consisting of 128~bits, 192 bits or 256 bits
respectively. We will describe the system when $|K|=2^{128}$ and
will indicate in Section~\ref{Sect:Relation} how to adapt the
algebraic description to the other situations.  For the Rijndael
algorithm we define
$$
K=M=C=R.
$$

Crucial for the description will be the following polynomial:
\begin{multline}                              \label{S-Box}
  \varphi(u):=\left({z}^{2}+1\right){u}^{254}+\left( {z}^{3}
    +1\right){u}^{253}+ \left( {z}^{7}+{z}^{6}+{z}^{5} +{z}^{4}
    +{z}^{3} +1\right){u}^{251}\\
  +\left ({z}^{5}+{z}^{2}+1\right){u}^{247}+\left( {z}^{7}
    +{z}^{6} +{z}^{5}+{z}^{4}+{z}^{2}\right){u}^{239}+{u}^{223}
  +\left ({z}^{7}+{z}^{5}+{ z}^{4}+{z}^{2}+1\right)
  {u}^{191}\\+\left ({z}^{7}+{z}^{3}+{z}^{2}+z+1 \right )
  {u}^{127}+(z^6+z^5+z+1)\in\F[u].
\end{multline}

Assume Alice and Bob share a common secret key $k\in R$ and Alice
wants to encrypt the message $m\in R$. In a first step both Alice
and Bob do a {\em key expansion} which will result in 10 elements
$k^{(t)}\in R\ t=0,\ldots,9$.

\paragraph{Key expansion:} Using the notation introduced in Equation\eqr{induced},
both Alice and Bob compute recursively 10 elements $k^{(t)}\in
R,\ t=0,\ldots,9$ in the following way:

\begin{eqnarray*}
  k^{(0)}& := & k\\
  k^{(t+1)}_0& := & \left(\sum_{i=0}^3
    \varphi(k^{(t)}_{i,3})x^i\right) x^3 + z^t + k^{(t)}_0\mbox{ for }t=0,\ldots,9.\\
  k^{(t+1)}_i& := &  k^{(t+1)}_{i-1} +  k^{(t)}_i \mbox{ for
    }t=0,\ldots,9,\ i=1,2,3.\\
\end{eqnarray*}

In order to describe the actual encryption algorithm we define
the ring element:
$$
\gamma := (z+1)x^3+x^2+x+z\in R.
$$

\paragraph{Rijndael encryption algorithm:} 
Using the round keys $k^{(t)}\in R$ and starting with the message
$m\in R$ Alice computes recursively:

\begin{eqnarray*}
  m^{(0)}& := & m+ k^{(0)}\\
  m^{(t+1)}& := &  \gamma\sum_{i=0}^3\sum_{j=0}^3 \varphi(m_{i,j}^{(t)})x^iy^{3i+j}+k^{(t+1)}
  \ \  \mbox{ for }t=0,\ldots,8.\\
  c:=m^{(10)}& := &   \sum_{i=0}^3\sum_{j=0}^3 \varphi(m_{i,j}^{(9)})x^iy^{3i+j}+k^{(10)} 
\end{eqnarray*}
The cipher to be transmitted by Alice is $c$. Note that in the
10th round no multiplication by $\gamma$ happens. This will make
sure that the decryption process follows formally the same
algebraic process as we will show next.

\paragraph{Rijndael decryption algorithm:} 

The polynomial $\varphi$ introduced in\eqr{S-Box} is a {\em
  permutation polynomial} describing a permutation of the
elements of $\F$. See Sections~\ref{Sect:Relation},~\ref{Sect:SBox} for more details.

There is a unique permutation polynomial $\psi(u)\in\F[u]$ of
degree at most $255$ such that
$\varphi\circ\psi=\psi\circ\varphi=id_{\F}$ and we will derive
this polynomial in Section~\ref{Sect:SBox}. The element
$\gamma\in R$ is invertible with
$$
\gamma^{-1} := (z^3+z+1)x^3+(z^3+z^2+1)x^2+(z^3+1)x+(z^3+z^2+z)\in R.
$$
Using the map $\psi$, the element $\gamma^{-1}$ and the round
keys $k^{(t)}$ Bob can decipher the message $m$ of Alice through:
\begin{eqnarray*}
  c^{(0)}& := & c+ k^{(10)}\\
  c^{(t+1)}& := &  \gamma^{-1}\sum_{i=0}^3\sum_{j=0}^3 
\psi(c_{i,j}^{(t)})x^iy^{i+j}+\gamma^{-1}k^{(9-t)}
  \ \  \mbox{ for }t=0,\ldots,8.\\
  c^{(10)}& := &   \sum_{i=0}^3\sum_{j=0}^3 
\psi(c_{i,j}^{(9)})x^iy^{i+j}+k^{(0)} 
\end{eqnarray*}

One readily verifies that $m=c^{(10)}$. Note that formally both
the encryption schedule and the decryption schedule follow the
same sequence of transformations. $\varphi$ is simply replaced by
$\psi$, multiplication by $\gamma$ is substituted with
multiplication by $\gamma^{-1}$ and the key schedule is changed
replacing $k^{(t)},t=0,\ldots,10$ with $k^{(10)},
\gamma^{-1}k^{(9)},\ldots,\gamma^{-1}k^{(1)},k^{(0)}$.

\begin{rem}
  Both encryption and decryption can be done very efficiently. In
  practice the polynomials $\varphi$ and $\psi$ are not evaluated
  and a look up table describing the permutations
  $\varphi,\psi:\F\longrightarrow\F$ is used instead.
  Substituting exponents $x^iy^j\mapsto x^iy^{3i+j}$ does not
  require any arithmetic and adding a round key $k^{(t+1)}$ is
  efficiently done through Boolean XOR operations. Arithmetic
  computations are required when multiplying by $\gamma$
  respectively by $\gamma^{-1}$.  Since in general multiplication
  by $\gamma$ is slightly easier than multiplication by
  $\gamma^{-1}$ the decryption algorithm takes in general
  slightly longer than the encryption algorithm.
\end{rem}

\begin{rem}
  (Compare with~\cite[page 55]{da02b} and~\cite{le02p}).
  $\gamma$ was chosen such that multiplication by $\gamma$ can be
  done with a minimal {\em branch number} and in the same time a
  good diffusion of $\F[x]/<x^4+1>$ is guaranteed. We are not
  convinced that the choice of $\gamma$ was optimal for the
  latter as it has a very small order in~$R$. A direct
  computation shows that $\gamma$ has order 4. 
  With this we also have an easy expression for $\gamma^{-1}$:
  \begin{equation}              \label{fac-gam}
  \gamma^{-1}=\gamma^3 = \gamma^2\gamma= (z^2x^2+z^2+1)\gamma.
  \end{equation}
  Instead of multiplying by $\gamma^{-1}$ it is therefore
  possible to multiply three times by $\gamma$ or alternatively
  one can pre-process the multiplication of $\gamma$ by the
  multiplication of $(z^2x^2+z^2+1)$. This is more efficient than
  multiplying the full expression by  $\gamma^{-1}$.
\end{rem}

\begin{rem}
We made a computer search for interesting factorizations of $
\gamma^{-1}$. It seems that the factorization\eqr{fac-gam} is
probably the easiest for computation purposes. The following is a
related interesting factorization which we found:
\begin{eqnarray}
  \gamma^{-1}&=& (zx^3+z+1)(x^3+(z^2+1)x^2+x+z^2)
\end{eqnarray}
\end{rem}

\Section{Relation to the Standard Description}
\label{Sect:Relation}

In the original description of the Rijndael algorithm the ring
$R$ was not used. Instead sets of elements having 128 bits were
described by a $4\times 4$ array each containing one byte, i.e.
$8$ bits. In order to relate the descriptions assign to each
element $r=\sum_{i=0}^3\sum_{j=0}^3 r_{i,j}x^iy^j$ the $4\times
4$ array
$$
\begin{array}{|c|c|c|c|}  \hline
r_{0,0}&  r_{0,1}& r_{0,2}&  r_{0,3}\\ \hline
r_{1,0}&  r_{1,1}& r_{1,2}&  r_{1,3}\\ \hline
r_{2,0}&  r_{2,1}& r_{2,2}&  r_{2,3}\\ \hline
r_{3,0}&  r_{3,1}& r_{3,2}&  r_{3,3}\\ \hline
\end{array}
$$
where each element $r_{i,j}\in\F$ is viewed as one byte.
Using a specific schedule the following operations are applied:

\paragraph{S-Box Transformation:} In this operation each element
$r_{i,j}\in\F$ is changed using a permutation $\varphi$ of the
symmetric group of $256$ elements. The permutation $\varphi$
decomposes into three permutations:
\begin{eqnarray}
\varphi_1:\  \F \longrightarrow \F, &&  f\longmapsto
\left\{\begin{array}{ccc}
f^{-1}&\mbox{if}& f\neq 0,\\ 
0&\mbox{ if } &f=0.\end{array}\right.\\
L: \ \F \longrightarrow \F, &&  f\longmapsto
(z^4+z^3+z^2+z+1)f\mod z^8+1.   \label{Lmap}\\ 
\varphi_3: \ \F \longrightarrow \F, &&  f\longmapsto
z^6+z^5+z+1+f.   \label{affine}
\end{eqnarray}
The permutation $\varphi$ is defined as $\varphi:=\varphi_3\circ
L\circ\varphi_1$. It is possible to describe the permutation
$\varphi$ using a permutation polynomial. For this note
that any permutation of $\F$ can also be described through a
unique interpolation polynomial (an element of $\F[u]$) having
degree at most $255$. We will denote this unique polynomial
describing the permutation $\varphi$ with $\varphi(u)$. The
context will always make it clear if we view $\varphi$ as a
permutation or as a polynomial $\varphi(u)\in\F[u]$.

This unique permutation polynomial can be computed in
the following way. If $\alpha\neq 0$ then
$$
T_\alpha(u):=u\sum_{i=0}^{254}\alpha^iu^{254-i}
$$
is the unique Lagrange interpolant having the property that
$$
T_\alpha(\beta)=\left\{\begin{array}{ccc}
1&\mbox{if}& \alpha=\beta,\\ 0&&\mbox{otherwise.}
\end{array}
\right.
$$
If $\alpha= 0$ then $T_\alpha(u)=u^{255}+1$ is the unique
Lagrange interpolant. The unique polynomial $\varphi(u)\in\F[u]$
is then readily computed using a symbolic algebra program as
$\varphi(u)=\sum_{\alpha\in\F}\varphi(\alpha)T_\alpha(u).$ This
computation was already done by Daemen and Rijmen in their
original proposal and the polynomial $\varphi$ can be found
in~\cite[Subsection 8.5.]{da99}

\paragraph{The ShiftRow Transformation:}

In this operation the bytes of the $i$th row are cyclically
shifted by $i$ positions. Algebraically this operation has a
simple interpretation. For this consider an element $r=r(x,y)\in
R$ as described in\eqr{induced}. The ShiftRow corresponds then
simply to the transformation:
$$
r=r(x,y)\longmapsto r(xy^3,y).
$$
This then translates in the encryption algorithm to replace
the monom $x^iy^j$ with the monom $x^iy^{3i+j}.$ The inverse of
the ShiftRow transformation is $r=r(x,y)\longmapsto r(xy,y)$
which translates into the replacement of $x^iy^j$ with the monom
$x^iy^{i+j}.$

\paragraph{The MixColumn Transformation:}

In this transformation each column $r_j=\sum_{i=0}^3
r_{i,j}x^i$ is multiplied by the element $\gamma$.

\paragraph{Add Round Key:} In this step the $t$-th round key
$k^{(t)}$ is added. 

\bigskip

The schedule of operation is as follows: In the `zero round'
the round key $k^{(0)}$ is simply added. In rounds 1-9 do the
operations `S-Box', `ShiftRow', `MixColumn' and `Add Round
Key'. In the 10th round do only `S-Box', `ShiftRow' and `Add Round
Key'. We have given the algebraic description for this schedule.

\subsection{AES-192 and AES-256}

Until now we described Rijndael when the key size and the message
size have 128 bits. This system is referred to as AES-128.  In
the original description~\cite{da99} one had the possibility to
vary both the size of the message blocks and the size of the
secret keys.

In the adopted standard~\cite{ae01} the size of the message
blocks are always taken to be 128 bits. In AES-192 and in AES-256
the secret key size consists of 192 respectively 256 bits. In
order to run these presumably more secure algorithms it will be
necessary to change the key expansion schedule of the last
section. In AES-192 13 elements $k^{(t)}\in R,\ t=0,\ldots,12$
are computed from the original 192 bits and the Rijndael
algorithm runs over 12 rounds. In AES-256 15 elements $k^{(t)}\in
R,\ t=0,\ldots,14$ are computed from the original 192 bits and
the Rijndael algorithm runs over 14 rounds. Other than this there
seems to be no difference and details can be found in~\cite{ae01,da02b}.

\Section{The Structure of the S-Box}
\label{Sect:SBox}

Except for the transformation of the S-Box all transformations
are $\Z_2$ linear. An understanding of the S-Box is therefore
most crucial. Surprisingly the permutation polynomial
$\varphi(u)$ is very sparse and we explain in this section why
this is the case.

The permutation $\varphi$ is the composition of the maps
$\varphi_1$, $L$ and $\varphi_3$. We will describe the permutation
polynomial for each of them.

The permutation polynomial for the map $\varphi_1$ is simply
given by $\varphi_1(u)=u^{254}$. 

The permutation $L$ is a  $\Z_2$ linear map. Because of
this reason there is a unique {\em linearized polynomial}
(see~\cite[Chapter 3]{li94})
${\mathcal L}(u)=\sum_{i=0}^7 \lambda_iu^{2^i}\in \F[u]$ such that 
$$
{\mathcal L}(f)=L(f)
$$
for all $f\in \F$.  If $\alpha_1,\ldots, \alpha_8$ is a any
 basis of $\F$ over the prime field $\Z_2$ then it is possible to
 compute the coefficients $\lambda_0,\lambda_1,\ldots,\lambda_7$ through the linear
 equations:
$$
{\mathcal L}(\alpha_j)=\sum_{i=0}^7
\lambda_i\alpha_j^{2^i}=L(\alpha_j),\ j=1,\ldots,8.
$$

This system of linear equations can be solved explicitly.  For
this let $\beta_1,\ldots, \beta_8$ be the dual basis (see
e.g.~\cite[Chapter 3]{li94}) of $\alpha_1,\ldots, \alpha_8$
characterized through the requirement:
$$
\mbox{Tr}_{\F/\Z_2}(\alpha_i\beta_j)=
\left\{\begin{array}{ccc}
1&\mbox{if}& i=j,\\ 
0&\mbox{ if } &i\neq j.
\end{array}
\right.
$$

Introduce the matrices:
$$
A:=
\left(
  \begin{array}{ccccc}
 \alpha_1 & \alpha_1^2 & \alpha_1^{4}&\ldots &\alpha_1^{2^7}\\
\alpha_2 & \alpha_2^2 & \alpha_2^4 &\ldots &\alpha_2^{2^7}\\
\vdots & \vdots &  & &\vdots\\
 \alpha_8 & \alpha_8^2 & \alpha_8^{4}&\ldots &\alpha_8^{2^7}
  \end{array}
\right)
\
B:=
\left(
  \begin{array}{cccc}
 \beta_1 & \beta_2 & \ldots &\beta_8\\
\beta_1^2 & \beta_2^2 & \ldots &\beta_8^2\\
\beta_1^4 & \beta_2^4 & \ldots &\beta_8^4\\
\vdots & \vdots &  &\vdots\\
\beta_1^{2^7} & \beta_2^{2^7} & \ldots &\beta_8^{2^7}
  \end{array}
\right)
$$
Assuming that $\beta_1,\ldots, \beta_8$ is the dual basis of
$\alpha_1,\ldots, \alpha_8$ simply means that $AB=I_8$.  

Let $S$ be the change of basis transformation such that 
$$
\left(
  \begin{array}{c}
 \alpha_1\\  \alpha_2\\ \vdots \\ \alpha_8
  \end{array}
\right) =S 
\left(
  \begin{array}{c}
 1\\ z\\ \vdots \\ z^7
  \end{array}
\right)
$$
and consider the matrix 
$$
L:=
\left(
  \begin{array}{cccccccc}
 1&0&0&0&1&1&1&1\\
 1&1&0&0&0&1&1&1\\
 1&1&1&0&0&0&1&1\\
 1&1&1&1&0&0&0&1\\
 1&1&1&1&1&0&0&0\\
 0&1&1&1&1&1&0&0\\
 0&0&1&1&1&1&1&0\\
 0&0&0&1&1&1&1&1\\
  \end{array}
\right)
$$
which describes the linear map introduced in\eqr{Lmap} with respect to the
polynomial basis $1,z,z^2,\ldots, z^7$.
Then one has:

\begin{lem}
  The coefficients $\lambda_0,\lambda_1,\ldots,\lambda_7$ of the
  permutation polynomial ${\mathcal L}(u)$ are given as:
\begin{equation}
  \left(
  \begin{array}{c}
 \lambda_0\\  \lambda_1\\ \vdots \\ \lambda_7
  \end{array}
\right) =
BSL^tS^{-1}
 \left(
  \begin{array}{c}
 \alpha_1\\  \alpha_2\\ \vdots \\ \alpha_8
  \end{array}
\right).
\end{equation}
\end{lem}
\begin{proof}
$SL^tS^{-1}$ describes the change of basis of the linear map $L$ with
regard to the basis $\alpha_1,\ldots, \alpha_8$.
\end{proof}

In order to explicitly compute the coefficients
$\lambda_0,\lambda_1,\ldots,\lambda_7$  we can work with the
polynomial basis $1,z,z^2,\ldots,z^7$ (in which case
$S=I_8$). Alternatively we can work with a {\em normal
  basis}. We explain the computation for a normal basis. 
Let
$$
\alpha := z^5+1\in \F.
$$
One verifies e.g. with the computer program Maple that
$\alpha$ is a primitive of $\F$ and that $\{
\alpha_i:=\alpha^{2^{i-1}}\mid i=1,\ldots,8\}$ forms a normal
basis. Such bases are called primitive normal bases.
$\alpha$ is special in the sense that it is the first element of
$\F$ with respect to lexicographic order which is both a
primitive and the generator of a normal basis.
\begin{rem}
  The existence of primitive normal bases has been established by
  Lenstra and Schoof~\cite{le87} for every finite extension
  $\mbox{GF}(q^m)$ of a finite field $\mbox{GF}(q)$.  Probably
  the nicest possible basis a finite field can have is a
  primitive normal basis which is also {\em self-dual}. We
  verified by computer search that $\mbox{GF}(256)$ does not have
  a self-dual, primitive normal basis.
\end{rem}

The dual basis of $\{\alpha_1,\ldots,\alpha_8 \}$ is readily
computed using Maple as $\{ \beta_j:=\beta^{2^{j-1}}\mid
j=1,\ldots,8\}$, where $\beta=z^5+z^4+z^2+1$. It is a well known
fact that the dual basis of a normal basis is normal as well.
The change of basis transformation is computed in this case as:
$$
S=
\left(\begin {array}{cccccccc} 
1&0&0&0&0&1&0&0\\
1&0&1&1&0&1&1&0\\ 
0&1&1&0&1&0&0&1\\ 
1&0&1&0&1&0&0&1\\ 
0&0&0&0&1&0&0&1\\ 
1&0&0&0&0&0&0&1\\ 
1&1&0&1&1&0&0&1\\ 
0&0&1&0&0&0&1&1\end {array}\right).
$$
With this one readily computes:
\begin{equation}
  \left(
  \begin{array}{c}
 \lambda_0\\  \lambda_1\\ \vdots \\ \lambda_7
  \end{array}
\right) =
BSL^tS^{-1}
 \left(
  \begin{array}{c}
 \alpha\\  \alpha^2\\ \vdots \\ \alpha^{2^7}
  \end{array}
\right) =
 \left(
  \begin{array}{c}
{z}^{2}+1\\
{z}^{3}+1\\
{z}^{7}+{z}^{6}+{z}^{5}+{z}^{4}+{z}^{3}+1\\
{z}^{5}+{z}^{2}+1\\
{z}^{7}+{z}^{6}+{z}^{5}+{z}^{4}+{z}^{2}\\
1\\
{z}^{7}+{z}^{5}+{z}^{4}+{z}^{2}+1\\
{z}^{7}+{z}^{3}+{z}^{2}+z+1
 \end{array}
\right) .
\end{equation}
The elements $\lambda_i$ already agree with the non-constant
coefficients of $\varphi$ introduced in\eqr{S-Box} up to order.
In order to get the exact form we need a polynomial description
of the permutation $\varphi_3$ introduced in\eqr{affine}.
Clearly the linear polynomial $\varphi_3(u):= u+1+z+z^5+z^6\in
\F[u]$ interpolates the affine map $\varphi_3$.

Concatenating the three polynomial maps we get:
$$
\varphi(u)=  \varphi_3\circ {\mathcal
  L}\circ\varphi_1(u)=1+z+z^5+z^6+{\mathcal L}(u^{254})\mod u^{256}+u. 
$$
Note that ${\mathcal L}$ has at most 8 nonzero
coefficients. Reducing ${\mathcal L}(u^{254})$ by the relation
$u^{256}=u$ will not change this and this explains the sparsity
of the polynomial $\varphi(u)$.

The fact that the permutation polynomial $\varphi(u)$ is
sparse does not imply that the inverse polynomial $\psi(u)$ is
sparse. For this note that 
$$
\psi(u) = \varphi_1^{-1}\circ{\mathcal L}^{-1}\circ \varphi_3^{-1}(u)\mod u^{256}+u.
$$
As before the coefficients of the polynomial $\calL^{-1}(u)$ are
computed from:
\begin{equation}
BS(L^{-1})^tS^{-1}
 \left(
  \begin{array}{c}
 \alpha\\  \alpha^2\\ \vdots \\ \alpha^{2^7}
  \end{array}
\right) .
\end{equation}
Using Maple we find:
\begin{multline}  
  {\mathcal L}^{-1}(u)= \left
    ({z}^{6}+{z}^{5}+{z}^{3}+{z}^{2}+z\right){u}^{128}+\left(
    {z}^{ 7}+{z}^{6}+ {z}^{4}+{z}^{3}+z+1\right) {u}^{64}\\
  +\left({z}^{6}+ {z}^{4}+ {z}^{3}+1\right ){u}^{32} +\left
    ({z}^{6}+{z}^{5}+{z}^{4}+{z}^{3} \right ){u}^{16}\\
  +\left({z}^{6}+{z}^{4}+{z}^{3}+z\right ){u}^{8}
  + \left({z}^{6}+{z}^{5}+{z}^{4}+{z}^{3}+{z}^{2}+z+1\right){u}^{4}\\
  + \left({z}^{7}+{z}^{6}+{z}^{5}+{z}^{4}+{z}^{3}+{z}^{2}
    +z\right) {u}^{2 } +\left ({z}^{2}+1\right )u\in\F[u].
\end{multline}
  
Combining the result with the map  $\varphi_3^{-1}(u)$ one gets:
\begin{equation}
\rho(u):=\calL^{-1}\varphi_3^{-1}(u)
=\calL^{-1}(u+\varphi_3(0))
=\calL^{-1}(u)+\calL^{-1}(\varphi_3(0))
=\calL^{-1}(u) +z^2+1.
\end{equation}

A polynomial of the form $\rho(u)$ is sometimes called an affine
polynomial~\cite{li94} reflecting the fact that the map
$\calL^{-1}\varphi_3^{-1}$ is affine linear over $\Z_2$. 

Concatenating $\rho(u)$ with the polynomial
$\varphi_1^{-1}(u)=\varphi_1(u)=u^{254}$ results in a non-sparse
polynomial $\psi(u)=\rho(u)^{254}\mod u^{256}+u$. For
completeness we provide the result of the Maple computation. The
coefficients are expressed in terms of the primitive $\alpha=z^5+1$.

{\footnotesize
\begin{multline*}  
\psi(u)={\alpha}^{163}{u}^{254}+{\alpha}^{76}{u}^{253}+{\alpha}^{195}{u}^{252}
+{\alpha}^{186}{u}^{251}+{\alpha}^{234}{u}^{250}+{\alpha}^{194}{u}^{
249}+{\alpha}^{248}{u}^{248}+{\alpha}^{255}{u}^{247}\\
+{\alpha}^{196}{u}
^{246}+{\alpha}^{100}{u}^{245}
+{\alpha}^{216}{u}^{244}+{\alpha}^{212}{u}^{243}+{\alpha}^{47}{u}^{242}
+{\alpha}^{17}{u}^{241}+{\alpha}^{85}{u}^{240}+{\alpha}^{103}{u}^{239}+
{\alpha}^{201}{u}^{238}\\
+{\alpha}^{184}
{u}^{237}+{\alpha}^{235}{u}^{236}+{\alpha}^{215}{u}^{235}+{\alpha}^{
170}{u}^{234}+{\alpha}^{74}{u}^{233}+{\alpha}^{15}{u}^{232}+{\alpha}^{
2}{u}^{231}+{\alpha}^{185}{u}^{230}+{\alpha}^{89}{u}^{229}+{\alpha}^{
26}{u}^{228}\\
+{\alpha}^{231}{u}^{227}+{\alpha}^{137}{u}^{226}+{\alpha}^
{110}{u}^{225}+{\alpha}^{230}{u}^{224}+{\alpha}^{20}{u}^{223}+{\alpha}
^{126}{u}^{222}+{\alpha}^{35}{u}^{221}+{\alpha}^{117}{u}^{220}+{\alpha
}^{48}{u}^{219}+{\alpha}^{141}{u}^{218}\\
+{\alpha}^{56}{u}^{217}+{\alpha
}^{29}{u}^{216}+{\alpha}^{154}{u}^{215}+{\alpha}^{207}{u}^{214}+{
\alpha}^{175}{u}^{213}+{\alpha}^{253}{u}^{212}+{\alpha}^{147}{u}^{211}
+{\alpha}^{5}{u}^{210}+{\alpha}^{43}{u}^{209}+{\alpha}^{194}{u}^{208}\\
+
{\alpha}^{242}{u}^{207}+{\alpha}^{202}{u}^{206}+{\alpha}^{27}{u}^{205}
+{\alpha}^{15}{u}^{204}+{\alpha}^{164}{u}^{203}+{\alpha}^{11}{u}^{202}
+{\alpha}^{233}{u}^{201}+{\alpha}^{56}{u}^{200}+{\alpha}^{121}{u}^{199
}+{\alpha}^{163}{u}^{198}\\
+{\alpha}^{69}{u}^{197}+{\alpha}^{113}{u}^{
196}+{\alpha}^{235}{u}^{195}+{\alpha}^{225}{u}^{194}+{\alpha}^{152}{u}
^{193}+{\alpha}^{227}{u}^{192}+{\alpha}^{9}{u}^{191}+{\alpha}^{78}{u}^
{190}+{\alpha}^{234}{u}^{189}+{\alpha}^{57}{u}^{188}\\
+{\alpha}^{136}{u}
^{187}+{\alpha}^{115}{u}^{186}+{\alpha}^{128}{u}^{185}+{\alpha}^{57}{u
}^{184}+{\alpha}^{223}{u}^{183}+{\alpha}^{228}{u}^{182}+{\alpha}^{110}
{u}^{181}+{\alpha}^{249}{u}^{180}+{\alpha}^{83}{u}^{179}+{\alpha}^{55}
{u}^{178}\\
+{\alpha}^{55}{u}^{177}+{\alpha}^{32}{u}^{176}+{\alpha}^{94}{
u}^{175}+{\alpha}^{71}{u}^{174}+{\alpha}^{88}{u}^{173}+{\alpha}^{94}{u
}^{172}+{\alpha}^{45}{u}^{171}+{\alpha}^{218}{u}^{170}+{\alpha}^{157}{
u}^{169}+{\alpha}^{73}{u}^{168}\\
+{\alpha}^{209}{u}^{167}+{\alpha}^{21}{
u}^{166}+{\alpha}^{122}{u}^{165}+{\alpha}^{127}{u}^{164}+{\alpha}^{206
}{u}^{163}+{\alpha}^{19}{u}^{162}+{\alpha}^{189}{u}^{161}+{\alpha}^{89
}{u}^{160}+{\alpha}^{177}{u}^{159}+{\alpha}^{192}{u}^{158}\\
+{\alpha}^{
211}{u}^{157}+{\alpha}^{99}{u}^{156}+{\alpha}^{195}{u}^{155}+{\alpha}^
{14}{u}^{154}+{\alpha}^{172}{u}^{153}+{\alpha}^{67}{u}^{152}+{\alpha}^
{136}{u}^{151}+{\alpha}^{6}{u}^{150}+{\alpha}^{122}{u}^{149}+{\alpha}^
{102}{u}^{148}\\
+{\alpha}^{198}{u}^{147}+{\alpha}^{14}{u}^{146}+{\alpha}
^{130}{u}^{145}+{\alpha}^{102}{u}^{144}+{\alpha}^{129}{u}^{143}+{
\alpha}^{246}{u}^{142}+{\alpha}^{187}{u}^{141}+{\alpha}^{85}{u}^{140}+
{\alpha}^{181}{u}^{139}+{\alpha}^{169}{u}^{138}\\
+{\alpha}^{230}{u}^{137
}+{\alpha}^{21}{u}^{136}+{\alpha}^{234}{u}^{135}+{\alpha}^{138}{u}^{
134}+{\alpha}^{104}{u}^{133}+{\alpha}^{26}{u}^{132}+{\alpha}^{229}{u}^
{131}+{\alpha}^{177}{u}^{130}+{\alpha}^{168}{u}^{129}+{\alpha}^{245}{u
}^{128}\\
+{\alpha}^{13}{u}^{127}+{\alpha}^{142}{u}^{126}+{\alpha}^{96}{u
}^{125}+{\alpha}^{240}{u}^{124}+{\alpha}^{224}{u}^{123}+{\alpha}^{32}{
u}^{122}+{\alpha}^{228}{u}^{121}+{\alpha}^{68}{u}^{120}+{\alpha}^{125}
{u}^{119}+{\alpha}^{147}{u}^{118}\\
+{\alpha}^{19}{u}^{117}+{\alpha}^{78}
{u}^{116}+{\alpha}^{51}{u}^{115}+{\alpha}^{114}{u}^{114}+{\alpha}^{87}
{u}^{113}+{\alpha}^{120}{u}^{112}+{\alpha}^{5}{u}^{111}+{\alpha}^{209}
{u}^{110}+{\alpha}^{51}{u}^{109}+{\alpha}^{39}{u}^{108}\\
+{\alpha}^{47}{
u}^{107}+{\alpha}^{109}{u}^{106}+{\alpha}^{159}{u}^{105}+{\alpha}^{203
}{u}^{104}+{\alpha}^{202}{u}^{103}+{\alpha}^{9}{u}^{102}+{\alpha}^{238
}{u}^{101}+{\alpha}^{44}{u}^{100}+{\alpha}^{188}{u}^{99}+{\alpha}^{234
}{u}^{98}\\
+{\alpha}^{59}{u}^{97}+{\alpha}^{15}{u}^{96}+{\alpha}^{131}{u
}^{95}+{\alpha}^{173}{u}^{94}+{\alpha}^{135}{u}^{93}+{\alpha}^{244}{u}
^{92}+{\alpha}^{216}{u}^{91}+{\alpha}^{50}{u}^{90}+{\alpha}^{218}{u}^{
89}+{\alpha}^{250}{u}^{88}
+{\alpha}^{108}{u}^{87}\\
+{\alpha}^{192}{u}^{86}+{\alpha}^{45}{u}^{85}+{\alpha}^{53}{u}^{84}
+{\alpha}^{186}{u}^{83}
+{\alpha}^{92}{u}^{82}+{\alpha}^{74}{u}^{81}+{\alpha}^{157}{u}^{80}+{
\alpha}^{172}{u}^{79}+{\alpha}^{99}{u}^{78}
+{\alpha}^{209}{u}^{77}+{
\alpha}^{236}{u}^{76}\\
+{\alpha}^{212}{u}^{75}+{\alpha}^{44}{u}^{74}+{
\alpha}^{209}{u}^{73}+{\alpha}^{175}{u}^{72}+{\alpha}^{101}{u}^{71}+{
\alpha}^{41}{u}^{70}+{\alpha}^{51}{u}^{69}+{\alpha}^{163}{u}^{68}
+{\alpha}^{183}{u}^{67}+{\alpha}^{245}{u}^{66}+{\alpha}^{169}{u}^{65}\\
+{\alpha}^{58}{u}^{64}+{\alpha}^{5}{u}^{63}+{\alpha}^{68}{u}^{62}+{
\alpha}^{63}{u}^{61}+{\alpha}^{202}{u}^{60}+{\alpha}^{138}{u}^{59}+{
\alpha}^{204}{u}^{58}+{\alpha}^{109}{u}^{57}+{\alpha}^{173}{u}^{56}+{
\alpha}^{214}{u}^{55}+{\alpha}^{61}{u}^{54}\\
+{\alpha}^{255}{u}^{53}+{
\alpha}^{185}{u}^{52}+{\alpha}^{249}{u}^{51}+{\alpha}^{153}{u}^{50}+{
\alpha}^{143}{u}^{49}+{\alpha}^{206}{u}^{48}
+{\alpha}^{163}{u}^{47}+{
\alpha}^{43}{u}^{46}+{\alpha}^{202}{u}^{45}+{\alpha}^{156}{u}^{44}+{
\alpha}^{70}{u}^{43}\\
+{\alpha}^{2}{u}^{42}+{\alpha}^{45}{u}^{41}+{
\alpha}^{81}{u}^{40}+{\alpha}^{43}{u}^{39}+{\alpha}^{121}{u}^{38}+{
\alpha}^{90}{u}^{37}+{\alpha}^{101}{u}^{36}+{\alpha}^{252}{u}^{35}+{
\alpha}^{42}{u}^{34}+{\alpha}^{176}{u}^{33}+{\alpha}^{201}{u}^{32}\\
+{\alpha}^{22}{u}^{31}+{\alpha}^{135}{u}^{30}+{\alpha}^{250}{u}^{29}+{
\alpha}^{176}{u}^{28}+{\alpha}^{76}{u}^{27}+{\alpha}^{90}{u}^{26}+{
\alpha}^{247}{u}^{25}+{\alpha}^{220}{u}^{24}+{\alpha}^{123}{u}^{23}+{
\alpha}^{76}{u}^{22}+\alpha\,{u}^{21}\\
+{\alpha}^{180}{u}^{20}+{\alpha}^
{108}{u}^{19}+{\alpha}^{222}{u}^{18}+{\alpha}^{54}{u}^{17}+{\alpha}^{
46}{u}^{16}+{\alpha}^{89}{u}^{15}+{\alpha}^{240}{u}^{14}+{\alpha}^{235
}{u}^{13}+{\alpha}^{208}{u}^{12}+{\alpha}^{194}{u}^{11}+{\alpha}^{2}{u
}^{10}\\
+{\alpha}^{201}{u}^{9}+{\alpha}^{67}{u}^{8}+{\alpha}^{247}{u}^{7
}+{\alpha}^{56}{u}^{6}+{\alpha}^{132}{u}^{5}+{\alpha}^{16}{u}^{4}+{
\alpha}^{242}{u}^{3}+{\alpha}^{223}{u}^{2}+{\alpha}^{243}u+{\alpha}^{
92}
\end{multline*}  }

Other than the fact that $\psi(u)=\rho(u)^{254}\mod u^{256}+u$
the author did not observe some regularity in the coefficients of
$\psi(u)$. The complicated algebraic structure of the inverse
S-Box shows that an algebraic attack on Rijndael which tries to
recursively solve the decryption equations might be very hard
indeed. Since $\varphi(u)$ is much more sparse it might be more
feasible to derive algebraic expressions of several rounds of the
encryption schedule.

Ferguson, Schroeppel and Whiting~\cite{fe01} show a way to
describe multiple rounds of the Rijndael algorithm using some
continued fraction expansion. The derived formulas look very
appealing. It is however not clear if there is any way to solve
these formulas by algebraic means.  Although algebraic
expressions for several rounds of Rijndael were derived it is our
belief that a compact polynomial description of several rounds
of Rijndael will result in an explosion of the variables. Further
research on this question will be needed.

In the last part of this section we provide the cycle
decomposition for the permutation of the S-Box. For this let
$\alpha=z^5+1$. We describe the cycles
$[\beta,\varphi(\beta),\varphi(\varphi(\beta)),\ldots]$ expressed
in terms of the primitive $\alpha$:

{\footnotesize
\begin{multline*}  \itemsep 20mm
  [\alpha, {\alpha}^{113}, {\alpha}^{139}, {\alpha}^{115},
  {\alpha}^{211}, { \alpha}^{233}, {\alpha}^{45}, {\alpha}^{150},
  {\alpha}^{25}, {\alpha}^{6}, {\alpha}^{96}, {\alpha}^{133},
  {\alpha}^{138}, {\alpha}^{80}, {\alpha}^{ 184}, {\alpha}^{130},
  {\alpha}^{119}, {\alpha}^{116}, {\alpha}^{222}, {\alpha}^{164},\\
  {\alpha}^{79}, {\alpha}^{114}, {\alpha}^{9}, {\alpha}^{165} ,
  {\alpha}^{160}, {\alpha}^{98}, {\alpha}^{81}, {\alpha}^{131},
  {\alpha}^{ 215}, {\alpha}^{181}, {\alpha}^{200},
  {\alpha}^{125}, {\alpha}^{143}, { \alpha}^{41}, {\alpha}^{179},
  {\alpha}^{202}, {\alpha}^{157}, {\alpha}^{70}, {\alpha}^{146},\\ 
  {\alpha}^{92}, 0, {\alpha}^{210}, {\alpha}^
  {232}, {\alpha}^{117}, {\alpha}^{11}, {\alpha}^{192},
  {\alpha}^{72}, { \alpha}^{185}, {\alpha}^{212}, {\alpha}^{21},
  {\alpha}^{105}, {\alpha}^{ 163}, {\alpha}^{216}, {\alpha}^{78},
  {\alpha}^{48}, {\alpha}^{174}, {\alpha }^{198}, {\alpha}^{209},
  {\alpha}^{176}, \alpha]
\end{multline*}  }\vspace{-1.2cm}

{\footnotesize
\begin{multline*}  
 [{\alpha}^{2}, {\alpha}^{112}, {\alpha}^{37}, {\alpha}^{161},
  {\alpha}^{ 242}, {\alpha}^{50}, {\alpha}^{240}, {\alpha}^{26},
  {\alpha}^{0}, {\alpha }^{42}, {\alpha}^{245}, {\alpha}^{168},
  {\alpha}^{10}, {\alpha}^{228}, { \alpha}^{229}, {\alpha}^{251},
  {\alpha}^{29}, {\alpha}^{76}, {\alpha}^{247}, {\alpha}^{223},
  {\alpha}^{243},\\ 
  {\alpha}^{17}, {\alpha}^{49}, {\alpha}^{197},
  {\alpha}^{225}, {\alpha}^{3}, {\alpha}^{104}, {\alpha}^{106},
  {\alpha}^{55}, {\alpha}^{32}, {\alpha}^{204}, {\alpha}^{203},
  {\alpha}^{132}, {\alpha}^{206}, {\alpha}^{19}, {\alpha}^{226},
  {\alpha}^{107}, {\alpha}^{84}, {\alpha}^{152}, {\alpha}^{231},
  {\alpha}^{142},\\ 
  {\alpha}^{159}, {\alpha}^{140}, {\alpha}^{110},
  {\alpha}^{162}, {\alpha}^{170}, {\alpha}^{248}, {\alpha}^{127},
  {\alpha}^{82}, {\alpha}^{148}, {\alpha}^{180}, {\alpha}^{151},
  {\alpha}^{31}, {\alpha}^{88}, {\alpha}^{227}, {\alpha}^{237},
  {\alpha}^{85}, {\alpha}^{43}, {\alpha}^{95}, {\alpha}^{218},
  {\alpha}^{71}, {\alpha}^{177}, \\{\alpha}^{121}, {\alpha}^{65},
  {\alpha}^{188}, {\alpha}^{186}, {\alpha}^{77}, {\alpha}^{23},
  {\alpha}^{187}, {\alpha}^{238}, {\alpha}^{167}, {\alpha}^{52},
  {\alpha}^{145}, {\alpha}^{136}, {\alpha}^{149}, {\alpha}^{147},
  {\alpha}^{123}, {\alpha}^{224}, {\alpha}^{20}, {\alpha}^{134},
  {\alpha}^{195}, {\alpha}^{2}]
\end{multline*}  }\vspace{-1.2cm}

{\footnotesize
\begin{multline*}  
  [{\alpha}^{4}, {\alpha}^{16}, {\alpha}^{69}, {\alpha}^{7},
  {\alpha}^{62}, { \alpha}^{34}, {\alpha}^{183}, {\alpha}^{172},
  {\alpha}^{208}, {\alpha}^{ 129}, {\alpha}^{220}, {\alpha}^{91},
  {\alpha}^{230}, {\alpha}^{153}, { \alpha}^{87}, {\alpha}^{102},
  {\alpha}^{234}, {\alpha}^{93}, {\alpha}^{51} , {\alpha}^{73},\\
  {\alpha}^{155}, {\alpha}^{196}, {\alpha}^{253}, {\alpha}^{124},
  {\alpha}^{101}, {\alpha}^{66}, {\alpha}^{235}, {\alpha}^{252},
  { \alpha}^{193}, {\alpha}^{18}, {\alpha}^{94}, {\alpha}^{90},
  {\alpha}^{144} , {\alpha}^{83}, {\alpha}^{5}, {\alpha}^{47},
  {\alpha}^{194}, {\alpha}^{244 }, {\alpha}^{118},\\
  {\alpha}^{173}, {\alpha}^{120}, {\alpha}^{199}, {\alpha}^{250},
  {\alpha}^{63}, {\alpha}^{156}, {\alpha}^{109}, {\alpha}^{221},
  { \alpha}^{30}, {\alpha}^{86}, {\alpha}^{46}, {\alpha}^{126},
  {\alpha}^{56}, {\alpha}^{44}, {\alpha}^{249}, {\alpha}^{33},
  {\alpha}^{24}, {\alpha}^{201 }, {\alpha}^{205}, {\alpha}^{191},\\
  {\alpha}^{128}, {\alpha}^{67}, {\alpha}^ {219}, {\alpha}^{239},
  {\alpha}^{15}, {\alpha}^{217}, {\alpha}^{103}, { \alpha}^{141},
  {\alpha}^{169}, {\alpha}^{241}, {\alpha}^{214}, {\alpha}^{ 59},
  {\alpha}^{154}, {\alpha}^{207}, {\alpha}^{175}, {\alpha}^{178},\\
  { \alpha}^{36}, {\alpha}^{97}, {\alpha}^{13}, {\alpha}^{28},
  {\alpha}^{12}, { \alpha}^{74}, {\alpha}^{182}, {\alpha}^{8},
  {\alpha}^{14}, {\alpha}^{58}, { \alpha}^{108}, {\alpha}^{75},
  {\alpha}^{4}]
\end{multline*}  }\vspace{-1.2cm}

{\footnotesize
\begin{multline*}  
  [{\alpha}^{22}, {\alpha}^{135}, {\alpha}^{64}, {\alpha}^{158},
  {\alpha}^{ 190}, {\alpha}^{189}, {\alpha}^{100}, {\alpha}^{40},
  {\alpha}^{60}, {\alpha }^{39}, {\alpha}^{99}, {\alpha}^{61},
  {\alpha}^{111}, {\alpha}^{166}, { \alpha}^{213}, {\alpha}^{27},\\
  {\alpha}^{89}, {\alpha}^{246}, {\alpha}^{171 }, {\alpha}^{137},
  {\alpha}^{122}, {\alpha}^{254}, {\alpha}^{35}, {\alpha}^ {57},
  {\alpha}^{53}, {\alpha}^{236}, {\alpha}^{68}, {\alpha}^{22}]
\end{multline*}  }\vspace{-1.2cm}

{\footnotesize
$$
[{\alpha}^{38}, {\alpha}^{54}, {\alpha}^{38}]\vspace{-1mm}
$$  }

It follows that $\varphi$ has cycle lengths 59, 81, 87, 27 and 2
and order 
$$
\lcm(59,81,87,27,2)=277,182
$$
confirming the result given by Lenstra~\cite{le02p}. We would
like to remark that the largest order an element of the symmetric
group of 256 elements can have is 451,129,701,092,070. In
comparison to this the order of $\varphi$ is not very large.

\Section{Conclusion}

In this paper we provided a description of the Advanced
Encryption Standard Rijndael which involved a series of 
polynomial transformations in a finite ring $R$. Special
attention was given to derive the permutation polynomials
describing the S-Box and the inverse S-Box of the Rijndael
system.


\begin{thebibliography}{10}

\bibitem{ae01}
Federal information processing standards publication 197, advanced encryption
  standard, November 2001.
\newblock Available at
  http://csrc.nist.gov/publications/fips/fips197/fips-197.pdf.

\bibitem{co98}
D.~Cox, J.~Little, and D.~O'Shea.
\newblock {\em Using Algebraic Geometry}.
\newblock Springer-Verlag, New York, 1998.

\bibitem{da99}
J.~Daemen and V.~Rijmen.
\newblock {\em {AES} Proposal {R}ijndael}, September 1999.
\newblock AES algorithm submission, available at
  http://csrc.nist.gov/encryption/aes/rijndael/.

\bibitem{da02b}
J.~Daemen and V.~Rijmen.
\newblock {\em The Design of {R}ijndael: {AES} -- The Advanced Encryption
  Standard}.
\newblock Springer-Verlag, Berlin Heidelberg, 2002.

\bibitem{fe01}
N.~Ferguson, R.~Schroeppel, and D.~Whiting.
\newblock A simple algebraic representation of {R}ijndael.
\newblock In A.M. Vaudenay, S.~Youssef, editor, {\em Selected Areas in
  Cryptography}, {LNCS} number 2259, pages 103--111. Springer Verlag, Berlin,
  December 2001.

\bibitem{le02p}
H.~W. {Lenstra, Jr.}
\newblock {R}ijndael for algebraists, April 2002.
\newblock Preprint: http://math.berkeley.edu/\~{}hwl/.

\bibitem{le87}
H.~W. {Lenstra, Jr.} and R.~J. Schoof.
\newblock Primitive normal bases for finite fields.
\newblock {\em Math. Comp.}, 48(177):217--231, 1987.

\bibitem{li94}
R.~Lidl and H.~Niederreiter.
\newblock {\em Introduction to Finite Fields and their Applications}.
\newblock Cambridge University Press, Cambridge, London, 1994.
\newblock Revised edition.

\bibitem{mu00}
S.~Murphy and M.~Robshaw.
\newblock New observations on {R}ijndael, August 2000.
\newblock Preprint: http://www.isg.rhul.ac.uk/\~{}mrobshaw/rijndael.pdf.

\bibitem{tr02b}
W.~Trappe and L.~C. Washington.
\newblock {\em Introduction to Cryptography with Coding Theory}.
\newblock Prentice Hall, Upper Saddle River, New Jersey, 2002.

\end{thebibliography}
\end{document}